\documentclass[twocolumn,notitlepage,showpacs,prl,superscriptaddress]{revtex4-1}

\usepackage{amsmath}
\usepackage{epsf}
\usepackage{graphicx}

\usepackage{dcolumn}
\usepackage{bm}

\usepackage{amssymb}
\usepackage{array}
\usepackage{varioref}
\usepackage{wrapfig}
\usepackage{etoolbox}
\usepackage{color}

\providecommand{\nn}{\nonumber}

\providecommand{\bv}[1]{\bm{\mathrm{#1}}}

\providecommand{\w}{\omega}
\providecommand{\W}{\Omega}
\providecommand{\q}{\bv{q}}
\providecommand{\Q}{\bv{Q}}
\providecommand{\p}{\bv{p}}

\renewcommand{\k}{\bv{k}}

\providecommand{\ve}{\varepsilon}

\providecommand{\vf}{v_F}
\providecommand{\vfv}{\bv{\vf}}
\providecommand{\kf}{k_F}

\providecommand{\pc}[1][ ]{\psi_{#1}^{\dagger}}
\providecommand{\pa}[1][ ]{\psi_{#1}}

\usepackage[caption=false]{subfig}

\providecommand{\kf}{k_F}

\renewcommand{\q}{\bv{q}}

\usepackage{etoolbox}

\begin{document}


\title{Critical behavior of itinerant fermions - role of finite size effects}

\author{Avraham Klein}
\email{ayklein@umn.edu}
\author{Andrey Chubukov}
\affiliation{School of Physics and Astronomy, University of Minnesota, Minneapolis. MN 55455}

\begin{abstract}
We study
 the role of finite size effects on
 a
 metallic critical behavior near a $q=0$ critical point
   and compare the results with the recent extensive quantum Monte-Carlo (QMC) study [Y. Schattner et al, PRX 6, 0231028]. This study found several features in both bosonic and fermionic responses, in disagreement with the expected critical behavior with dynamical exponent $z=3$. We show that finite size effects are particularly strong for $z=3$ criticality and give rise to a
     behavior different from that of an infinite system, over a wide range of momenta and frequencies. We argue that by taking
       finite size
       effects into account, the QMC results can be explained within $z=3$ theory. Our results also have implications for small interacting fermionic systems, such as magnetic nanoparticles.
\end{abstract}
\maketitle

{\bf {\it Introduction}}~~~
Critical behavior in itinerant fermionic systems is a fascinating subject, which has attracted much interest in recent years, with particular emphasis on the behavior in two dimensions (2D)\cite{Revh,Reve}. Near a 2D quantum critical point (QCP), soft bosonic fluctuations of the order parameter field mediate strong interaction between low-energy fermions and destroy Fermi-liquid (FL) behavior down to a progressively small energy $\omega_{FL}$, which vanishes at a QCP. Simultaneously, low-energy fermions affect soft bosonic fluctuations by (i) providing Landau damping and (ii) changing the bosonic mass
. The destruction of the FL holds even if the overall strength of the interaction is much smaller than the fermionic bandwidth, i.e. when fermions remain itinerant throughout the transition.

Before the feedback from low-energy fermions is included, the inverse propagator of a soft boson is generally assumed to be an analytic function of momentum and frequency: $\chi^{-1} (\q, \Omega_m) \propto (|\q-\Q|^2 + \Omega^2_m/c^2)$, where ${\bf Q}$ is the momentum at which order parameter fluctuations condense at a QCP and $\Omega_m$ are Matsubara frequencies. The Landau damping comes from the insertion of the fermionic particle-hole bubble into the bosonic propagator.
     The form of the Landau damping term depends on whether $\Q$ has a finite value (e.g. $(\pi,\pi)$ for a SDW QCP), or is zero, as for a nematic or a ferromagnetic QCP. In the first case the Landau damping term scales
      as just $|\Omega_m|$, while in the second case it scales as $|\Omega_m|/q$. In both cases, the Landau damping term wins at small $\Omega_m$ over the bare $\Omega^2_m$ and changes the dynamical exponent from $z=1$ to $z=2$ for $Q \neq 0$ and to $z=3$ for $Q=0$. The one-loop fermionic self-energy due to scattering by Landau overdamped critical fluctuations has a non-FL frequency dependence in 2D: $\Sigma (\omega_m) \propto \omega^{1-1/z}_m$ ($\omega^{1/2}$  at particular hot spots along the Fermi surface (FS), when $Q\neq 0$, and $\omega^{2/3}$ everywhere on the FS, when $Q=0$) \cite{Lee1993,Altshuler1994,Nayak,Millis,acs,patel,senthil}.

     For $z=2$, the forms of the fermionic and bosonic propagators in 2D are further affected by logarithmically singular higher-loop corrections from low-energy fermions~\cite{acs}, and the dynamical exponent $z$ likely flows away from $z=2$ \cite{Metlitski2010}.
       For $z=3$, the corrections are also logarithmically singular, but
        singularities show up only at
         three-loop and higher orders \cite{Lee2009,Metlitski2010a,Mross,Metzner2015,Mandal}.
     Because singular three-loop corrections have quite small prefactors, one could generally expect
      the $z=3$ scaling to remain valid down to the lowest
        frequencies (and, possibly, all frequencies~\cite{subir_last}). In particular, one could expect the $z=3$ behavior to be reproduced in numerical calculations, which probe the system at a finite $T$, when bosonic and fermionic Matsubara frequencies are discrete.  It was quite surprising in this respect
          that the
          recent Quantum Monte Carlo (QMC) analysis of a model, designed to emulate a 2D nematic transition~\cite{Schattner2016}, found seemingly $z=2$ behavior
           over a range of temperatures and frequencies.
     Furthermore, the same study found  that the quasiparticle residue $Z = 1/(1+ d\Sigma/d\omega)$ remains finite down to the lowest frequencies when tuning across the critical point. Such disagreement with a basic, established theory is intriguing and should be understood.

Several known mechanisms can make it difficult to extract $|\Omega|/q$ behavior from the data on $\chi (q, \Omega)$. First, when fermionic residue is small, $|\Omega|/q$ scaling is observed only when $v_F q > \Omega_m/Z$, a more severe restriction than just $v_F q > \Omega_m$ \cite{Chubukov_Maslov}. Second, because the nematic order is not a conserved quantity, the bosonic propagator in 2D has an additional $q-$independent $\Omega^{2/3}$ term~\cite{omega23}. This term does not break $z=3$ scaling but can mask $|\Omega|/q$  behavior. The
form of the bosonic propagator is further complicated at finite $T$
 because of special contributions from thermal fluctuations, which act much like impurities~\cite{DellAnna2006,Punk2016}.
  Third, if critical bosons are separate degrees of freedom, rather than collective modes of fermions, they may have their own damping in addition to Landau  damping, and that damping doesn't have to have $\Omega/q$ form. These mechanisms, particularly the last one, were essential to understand the violation of $\Omega/q$ scaling in uranium-based itinerant ferromagnets UGe$_2$ and UCoGe \cite{Huxley2003,Stock2011,Chubukov2014}. They are, however, less relevant to QMC analysis because in this analysis intrinsic dynamics of bosons (fluctuations of localized spins of the transferred Ising model) can be separated from the effects due to fermions by switching on and off the coupling between the two degrees of freedom.

 \begin{figure*}
  \centering
  \subfloat[\label{fig:pi-plot}]{
    \includegraphics[width=0.3\textwidth, clip,trim=0 180 0 100]{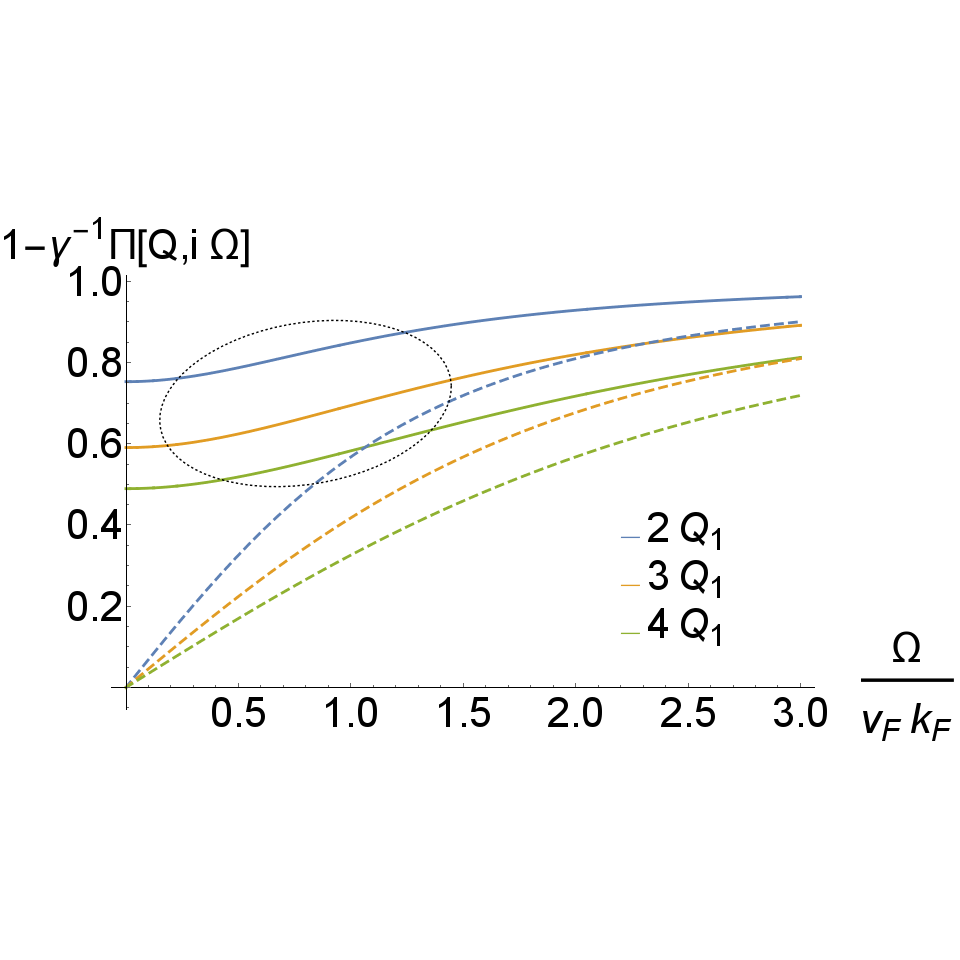}}
  \subfloat[\label{fig:data-plot}]{
    \includegraphics[width=0.3\textwidth, clip,trim=0 180 0 100]{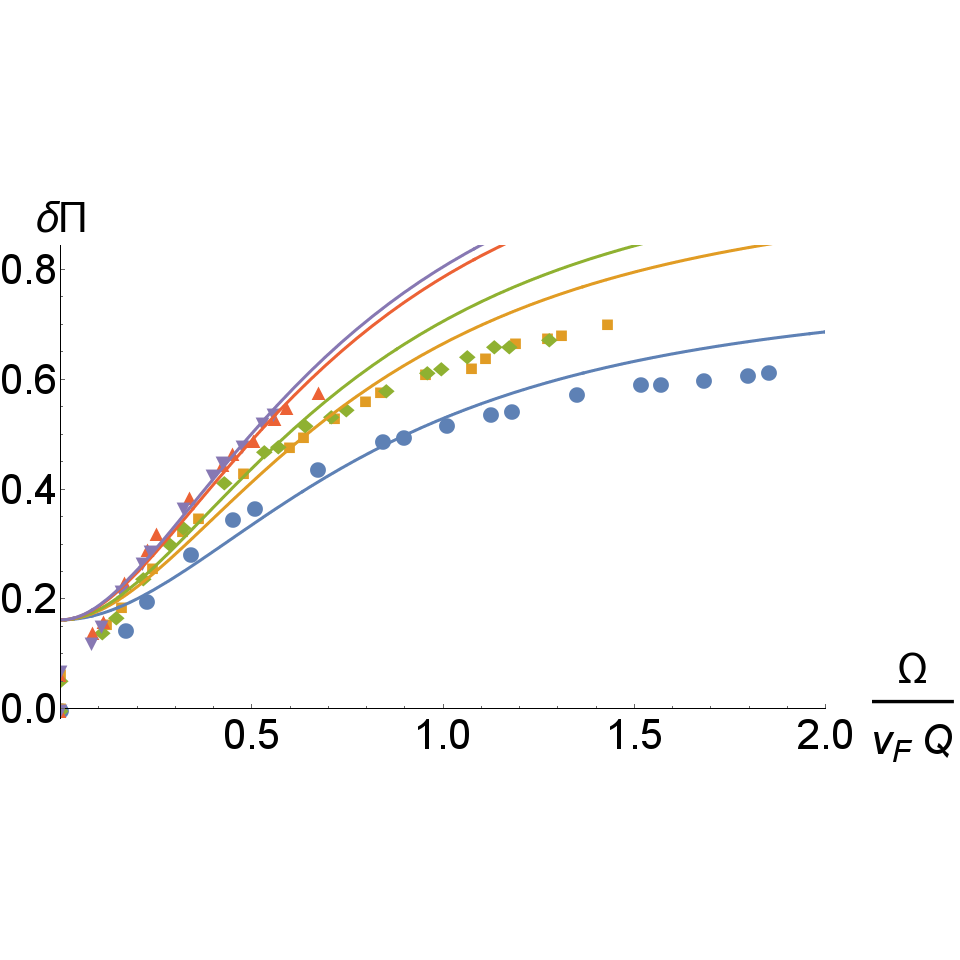}}
  \subfloat[\label{fig:scaling-plot}]{
    \includegraphics[width=0.3\textwidth, clip,trim=0 180 0 100]{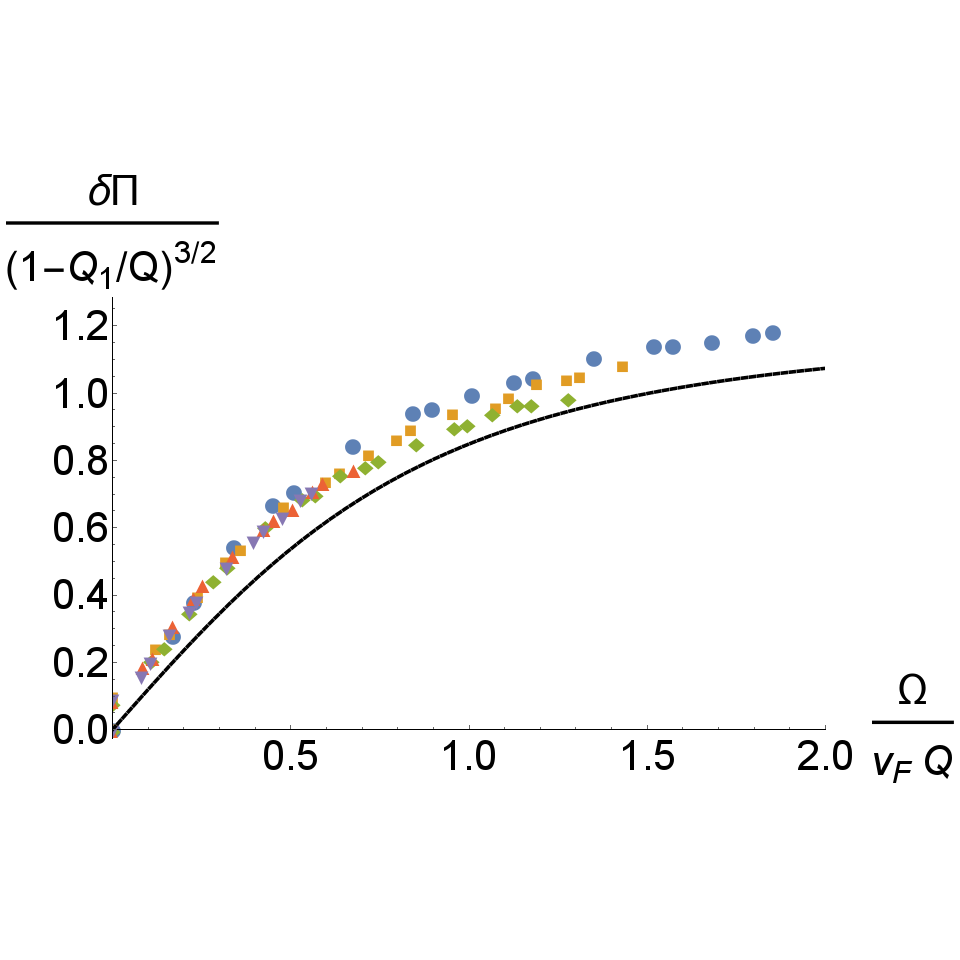}}

  \caption{(color online) Dynamic polarization bubble $\Pi(q, \Omega_m)$.  In an infinite system, $\Pi(q, \Omega_m) = \gamma (1 - |\Omega_m|/(v_F q)^2 + \Omega^2_m)$, i.e., the slope of $1- \Pi (q, \Omega_m)/\gamma$  scales as $1/q$. In a finite system, this behavior is modified because $\Pi(q, \Omega_m)$ vanishes
   at $q_1 =\pi / L$.  Panel (a) shows
  $1- \Pi (q, \Omega_m)/\gamma$ vs $\Omega_m$ at different $q$ in a finite square lattice of size $L = 20 a$ (solid lines) and an infinite system (dashed lines). For the finite system there is a large intermediate region (dotted ellipse), where the slope of $\Pi$ appears roughly independent of $q$.
    Panel (b) -- the frequency variation of $\Pi (q, \Omega_m)$  in our model vs QMC data from Ref. \cite{Schattner2016}. [26]
    Panel (c) -- phenomenological form of $\Pi (q, \Omega_m)$, Eq. (\ref{eq:Pi-phem}), vs QMC data. [26] }
  \label{fig:xi-fig}
\end{figure*}

In this work we explore an additional, hitherto
undiscussed aspect of the problem -- a strong sensitivity of an itinerant QC system to finite-size effects. To separate this from the effects associated with the non-conservation of the nematic order parameter, we approximate the nematic form-factor by a constant, i.e., equate nematic fluctuations with fermionic density fluctuations.
  We show that in a finite system of size $L$, the polarization bubble, whose dynamical part yields the Landau damping at $L \to \infty$, is
\begin{equation}
  \label{eq:damping-intro}
  \Pi (q, \Omega_m) = \Pi\left(\alpha,\beta\right), ~~\alpha = \frac{|\Omega_m|}{v_F q}, ~~\beta = \frac{q_1}{q}, \qquad  q_1 = \pi / L.
\end{equation}
When $q_1/q$ is vanishingly small,  $\Pi(\alpha,\beta) = \gamma (1-\alpha/\sqrt{1+\alpha^2})$, as for an infinite system.  At at a non-zero $\beta$, the form of $\Pi(\alpha,\beta)$ is determined by a combination of two effects: (i) $\Pi (0, \beta)$ decreases with increasing $\beta$ ($\Pi (\alpha,\beta)$ vanishes at $\beta =1$, see below), and (ii) $\Pi (\alpha, \beta)$ vanishes at $\alpha \to \infty$ for any $\beta$.  As a result, there appears an intermediate range of $\alpha < 1$, where the variation of $\Pi (\alpha,\beta)$ vs $\alpha$
(i.e., vs $|\Omega_m|$) is roughly linear, but the slope decreases as $\beta$ increases and over a rather wide range of parameters
appears almost independent of $q$
(see Fig. \ref{fig:pi-plot}).  This mimics $z=2$ scaling as reported in \cite{Schattner2016} (Fig. \ref{fig:data-plot}). We also found that the data from Ref. \cite{Schattner2016} can be reproduced in an alternative, semi-phenomenological approach, by invoking the fact that in a finite system the polarization bubble vanishes not at $q=0$, but at at $q=q_1$, i.e. at $\beta =1$. Near $\beta =1$, $\Pi (\alpha,\beta) \propto (1-\beta)^{3/2}$ (see below). Assuming
 phenomenologically
 that this is the main finite-size effect, we approximate the frequency dependence of the polarization bubble as
\begin{equation}
  \label{eq:Pi-phem}
  \Pi (\alpha, \beta) = \Pi (\alpha,0)(1-\beta)^{3/2} = \gamma \left(1-\frac{\alpha}{\sqrt{1+\alpha^2}}\right) (1-\beta)^{3/2}.
\end{equation}
This simple form reproduces the data from~\cite{Schattner2016} to surprisingly good accuracy (see Fig. \ref{fig:scaling-plot}).

 The fermionic self-energy also has strong finite-size dependence. We found (for $\omega_m >0$)
  \begin{equation}
  \label{eq:sigma-intro}
  \Sigma(\omega_m) \propto \omega_1^{2/3}\left[\left(1 + \frac{\omega_m}{\w_1}\right)^{2/3}-1\right]
\end{equation}
where $\w_1 \sim \vf q_1$
 up to logarithms (see Eq. (14) below).
At $\omega \gg \omega_1$ this yields $\Sigma (\omega_m) \propto \omega^{2/3}$, as in the infinite system. However, at smaller $\omega$, $\Sigma (\omega_m) = a \omega_m + b \omega^2_m + ...$ as in a FL. As a result, for probes at $\omega \geq \omega_1$, it looks as if the quasipartcle residue remains finite throughout the transition. We compared Eq. (\ref{eq:sigma-intro}) with Ref. \cite{Schattner2016} and again found good agreement with QMC data (see Fig. 3).

\begin{figure}[t]
  \centering
  \subfloat{
    \includegraphics[width=0.45\columnwidth]{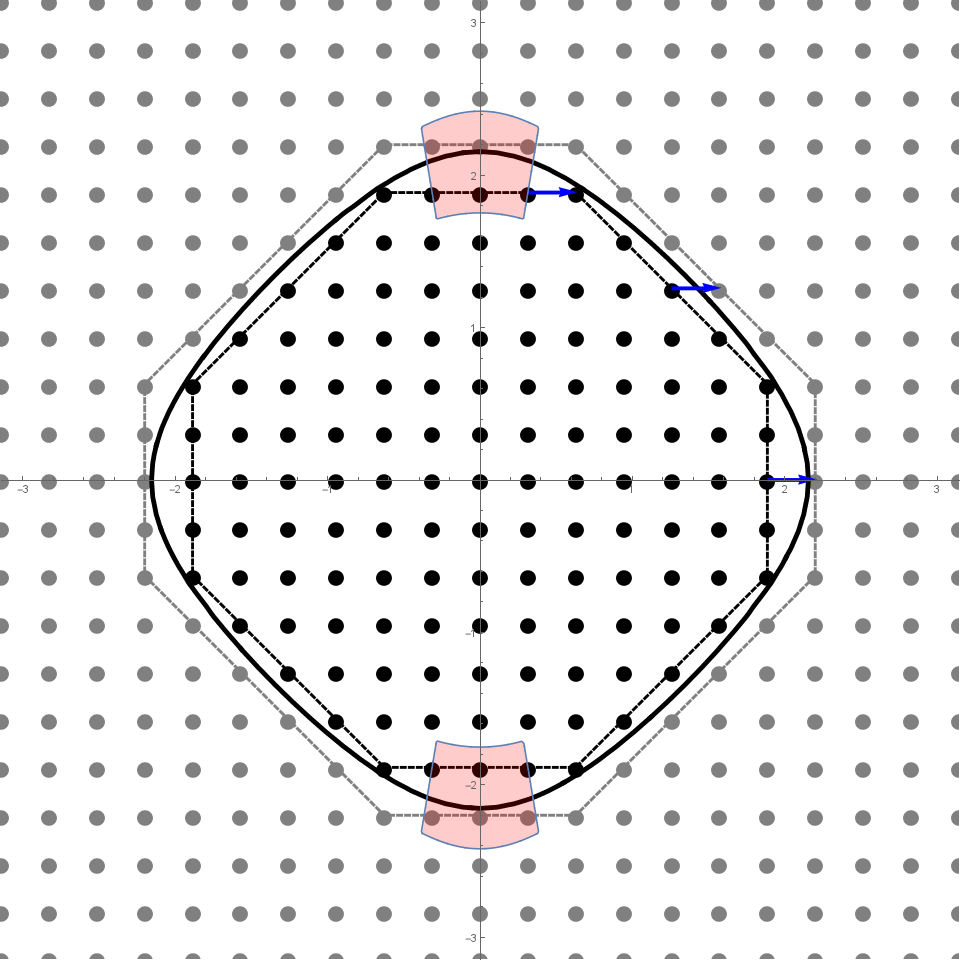}}
  ~
  \subfloat{
    \includegraphics[width=0.45\columnwidth]{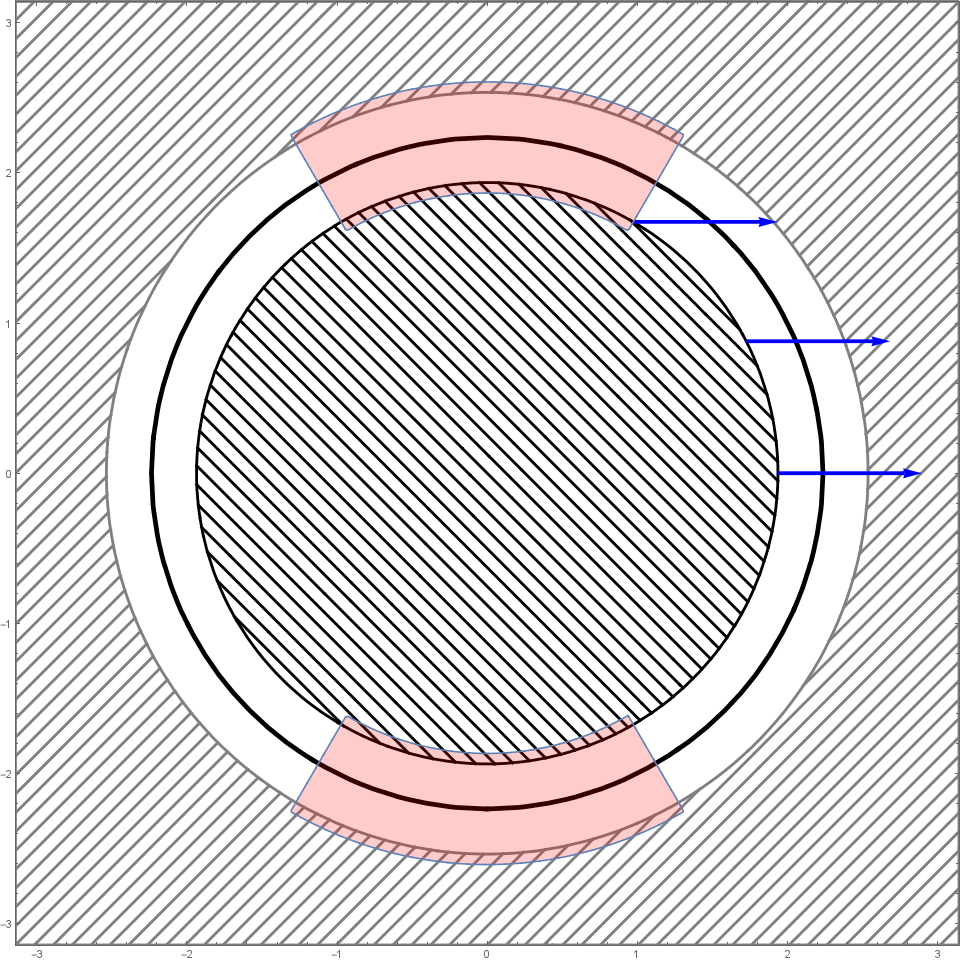}}
  \caption{Coupling of bosonic fluctuations and
   fermions
   in a finite system. The panel depicts the reciprocal space of a typical square lattice. The solid black line is the Fermi surface that would exist in an infinite system. Black (gray) dots are the filled (empty) states in $\k$-space. A
    bosonic fluctuation of wavevector $\q$ (blue arrows) can couple to
      fermions
      by exciting an electron-hole pair. In an infinite system, this coupling can occur at any point on the Fermi surface. However, in a finite system, the filled (empty) states actually appear as a series of facades (dashed lines). As a result, for small enough $q$ there is a region of the Fermi surface where excitations cannot occur (shaded region). The right panel depicts an approximation to the left panel, replacing the lattice with the electron/hole continuum
       with a  gap in momentum space with the
        width $q_1=\pi/L$, as described in Eqs. (\ref{eq:h-i-cont}) and (\ref{eq:cutoff-cont}).
  \label{fig:facades}}
\end{figure}

{\bf {\it Model calculations}}~~~
We consider a 2D system of size $L\times L$.
The system is composed of electrons
 hopping on a lattice and coupled to a scalar boson
 field.
The free propagators of electrons and bosons are of the form,
\begin{align}
  \label{eq:free-prop-elec}
  G^{-1}(\k,i\w) &= i\omega - \ve_{\k}, \\
  \chi^{-1}(\q,i\W) &= \chi_0^{-1}(m^2 + q^2 + \W^2/c^2) ,
\end{align}
where $v_0$ is the bosonic velocity, and $m = 1/\xi$ goes to zero at the QCP. In a finite system the interaction can be written as
\begin{equation}
  \label{eq:finite-int}
  H_I = -g \sum_{\q,\k} f_{\k} \phi_{\q}\pc[\k + \frac \q 2]\pa[ \k - \frac\q 2],
\end{equation}
where $f_{\k}$ is a form factor, the $\q,\k$ sums are over the 1st bosonic and fermionic BZ's respectively, and $g$ is a coupling constant. If we restrict our attention to states near the Fermi surface, we can rewrite the interaction as:
\begin{align}
  \label{eq:h-i-cont}
  H_I \propto \sum_{\q} \int d\theta d\epsilon_k ~ f(\theta)
  \pc[\k(\theta) - \frac \q 2]\pa[\k(\theta) + \frac\q 2] \Phi(\q,\epsilon_k,\theta).
\end{align}
Here, $\epsilon_k = \vf (k-\kf)$ measures the distance from the FS and $f(\theta)$ is the form-factor at the FS.
 We separate the effects due to order parameter non-conservation from finite-size effects, and focus on the latter, by setting $f(\theta)=1$. The function $\Phi$ is an indicator function due to finite size of a system. It accounts for the lack of $k-$states in an annulus of width $q_1$ around the FS (see Fig. \ref{fig:facades} for visualization),
\begin{align}
  \label{eq:cutoff-cont}
  \Phi(\q, \epsilon_k, \theta) &=
                               \Theta(|2\epsilon_k + \vfv(\theta)\cdot \q| -\vf q_1) \times \nn\\
                             &  \qquad\quad \Theta(|2\epsilon_k - \vfv(\theta)\cdot\q| - \vf q_1),
\end{align}
Let us recall the origin of the form $\Omega/\vf q$ for the polarization. It arises from the fact that for small enough $\W \ll \vf |\q|$ there is \emph{always} an electron-hole pair that can be resonantly excited, given by the condition $\W = \vfv\cdot\q = \vf q \cos\theta$, as long as the FS is closed. However, as Eq. \eqref{eq:h-i-cont} shows, in a finite system and for small enough $\q$ it is not always possible to find such a pair.
The reason for this is that in a finite system the border between filled and empty states is not a smooth curve but a series of ``facades'' (Fig. \ref{fig:facades}).
At small enough momentum and frequency it is no longer possible to find a resonant pair that also conserves momentum. This gives a lower cutoff of $\W < \vf q \sim \vf q_1$ for the overdamped behavior of the bosonic excitations, which introduces the new scale
$\beta = q_1/q$, leading to Eq. \eqref{eq:damping-intro}.
For $\beta$ close to one, the suppression effect is proportional to $(1-\beta)^{3/2}$. To see this, note
 that as we move around the FS, the available phase space for particle-hole excitations is $\vfv\cdot \q - \vf q_1 = \vf q(\cos\theta - \beta)$.
Because of this restriction,
 the polarization bubble
 is proportional to
\begin{equation}
  \label{eq:Pi-beta-leading}
  \int_0^{\theta_\beta}d\theta~(\cos\theta - \beta) \sim (1-\beta)^{3/2},\quad \theta_\beta = \cos^{-1}\beta,
\end{equation}
(we used
 $\theta_\beta \sim (1-\beta)^{1/2}$ for $1-\beta \ll 1$). This is the reasoning behind Eq. \eqref{eq:Pi-phem}.

In addition to
  the damping term, the polarization bubble has
    a
      static piece, which  renormalizes
      the bosonic mass and shifts
      the position of the QCP. 
       This last term also gets modified in a finite system in such  a way that
        the bosonic mass remains positive at a QCP of an infinite system, i.e., finite-size effects shift the system
       \emph{away} from the
         critical point.
 This finite bosonic mass affects the self energy.
   In an infinite system
    $\Sigma(\q,\w) \sim \w^{2/3}$ displays a non-FL behavior
   at a QCP. In a finite-size system,  the mass term protects the FL behavior at low frequencies, which is the content of Eq. \eqref{eq:sigma-intro}.

We demonstrate
this behavior
 by explicitly calculating the polarization bubble and
 the
  self energy for the approximate model of Eqs. \eqref{eq:free-prop-elec}-\eqref{eq:cutoff-cont}. We assume a parabolic dispersion and calculate the one-loop diagrams. To calculate the one-loop polarization bubble at $T\to 0$ we take into account only those $k-$ states that are on opposite sides of
   the boundary of
   the Fermi surface. In this case the indicator function can be recast as
\begin{align}
  \label{eq:cutoff-cont-2}
  \Phi &= \Theta\left(\left|\vfv(\theta)\cdot\q\right| - \vf q_1\right) \times \nn\\
  &\qquad\quad \Theta\left(\vfv(\theta)\cdot\q - \vf q_1 - \left|2\epsilon_k\right|\right)
\end{align}
The polarization is then given by:
\begin{equation}
  \label{eq:Pi-model}
  \Pi({\bf q},\W) = \frac{m_0 g^2\chi_0}{\pi^2}\int_{0}^{\theta_{\beta}}\frac{(\cos\theta-\beta)\cos\theta}{\alpha^2 + \cos^2\theta}
  ~d\theta.
\end{equation}
where $m_0$ is the bare fermionic mass. The limits of the integration are precisely those defined by the finite size effect of Eq. \eqref{eq:cutoff-cont-2}.
Evaluating the integrals we find
\begin{widetext}
\begin{equation}
  \label{eq:pi-1}
\Pi(\alpha,\beta) = \frac{2\gamma}{\pi} \left[ \cos^{-1}{\beta} - \frac{\alpha}{(1+\alpha^2)^{1/2}} \left(\tan^{-1}{\frac{\alpha}{\beta} \frac{(1-\beta^2)^{1/2}}{(1+\alpha^2)^{1/2}}} + \frac{\beta}{\alpha} \tanh^{-1}{\frac{(1-\beta^2)^{1/2}}{(1+\alpha^2)^{1/2}}}\right)\right]
\end{equation}
\end{widetext}
where $\gamma =m_0g^2\chi_0/2\pi$.  One can easily check that $\Pi (\alpha,\beta)$ vanishes at $\alpha \to \infty$ and at $\beta = 1$.
When $\beta \ll \alpha <1$,  $\Pi (\alpha,\beta) \approx \gamma\left(1 - \beta\log{2/\alpha} -\alpha\right)$
Near $\beta =1$, $\Pi (\alpha,\beta) \approx 1.2\gamma (1-\beta)^{3/2}/(1+\alpha^2)$.

 We next calculate the fermionic self-energy
\begin{align}
  \label{eq:sigma-def}
  \Sigma(\k, \w_m) &= g^2 \int \frac{d\W d^2q}{(2\pi)^3} \frac{\tilde{\chi}(\q,\W_m)} {i(\w_m+\W_m) - \vfv\cdot\q},
\end{align}
where
${\tilde \chi}^{-1} = \chi^{-1} + \Pi$. We used the form of $\Pi$ at
$\beta < \alpha \ll 1$
 and absorbed the constant
 term in $\Pi$
  into $m^2$.
    One can verify
    that the integral is dominated by $|\W_m|
    \sim \w_m, \hat{\k}_F\cdot \q \sim \w_m, \hat{\k}_F\times\q \sim \mbox{max}(|\Omega_m|,q_1\log2/\alpha)^{1/3}$. Thus, in the range
$  \W_m < \vf q_1 \log{\gamma^{1/2}/q_1}$
 we can
  expand in the dynamic part of the susceptibility, which makes
   $\Sigma$ an analytic function of frequency.  An evaluation of Eq. \eqref{eq:sigma-def} at
   $m=0$ yields
\begin{align}
  \label{eq:sigma-final}
  i\Sigma(\w_m) &= \frac{1}{\sqrt 3}\left(\frac{\gamma}{\vf}\right)^{2/3}(\vf q_1 \log \alpha_L)^{2/3}\times \nn \\ &\quad \times\left[\left(1+ \frac{\w_m}{\vf q_1 \log\alpha_L}\right)^{2/3}-1\right],
\end{align}
where
\begin{equation}
  \label{eq:alpha-L-def}
  \log\alpha_L = \frac{4}{3}\log\frac{\vf q_L}{\w_m},\quad q_L = (8\gamma q_1^2)^{1/4}.
\end{equation}

\begin{figure}
   \centering
   \subfloat[\label{fig:sig-pl}]{
     \includegraphics[width=0.24\textwidth, clip,trim=0 180 0 100]{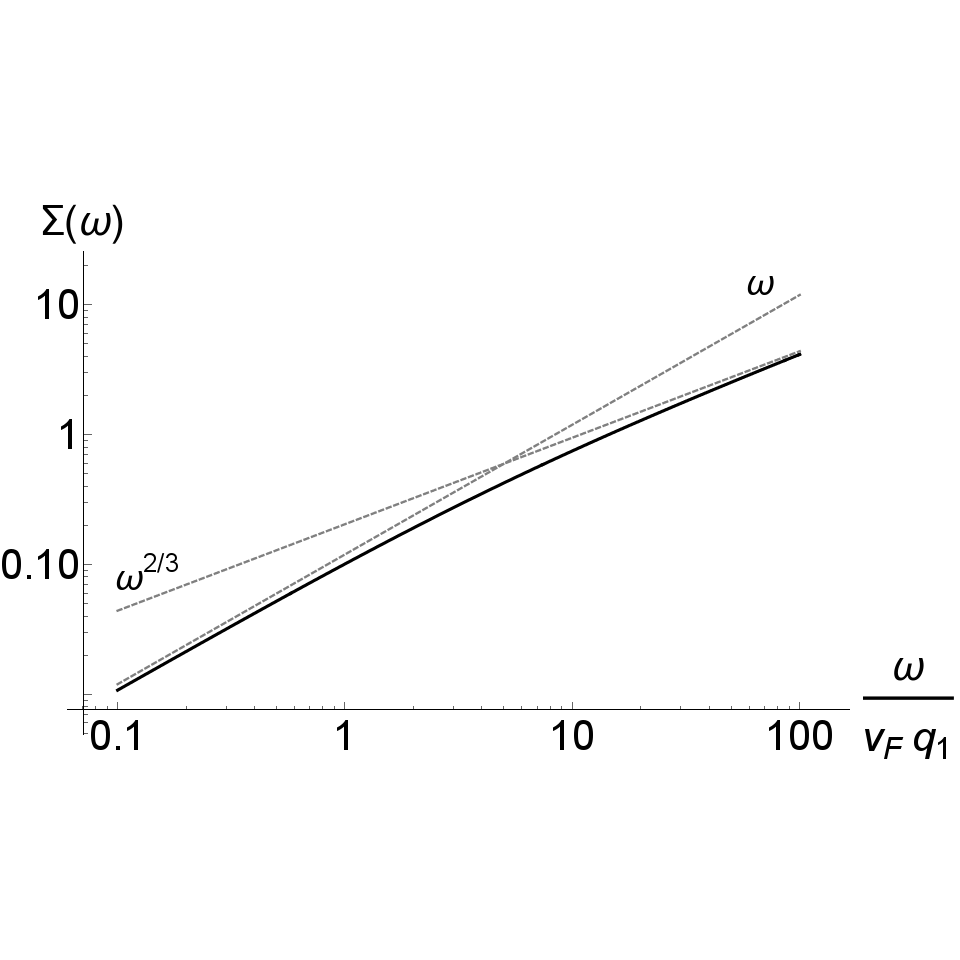}}
   \subfloat[\label{fig:zqp-pl}]{
     \includegraphics[width=0.24\textwidth, clip,trim=0 180 0 100]{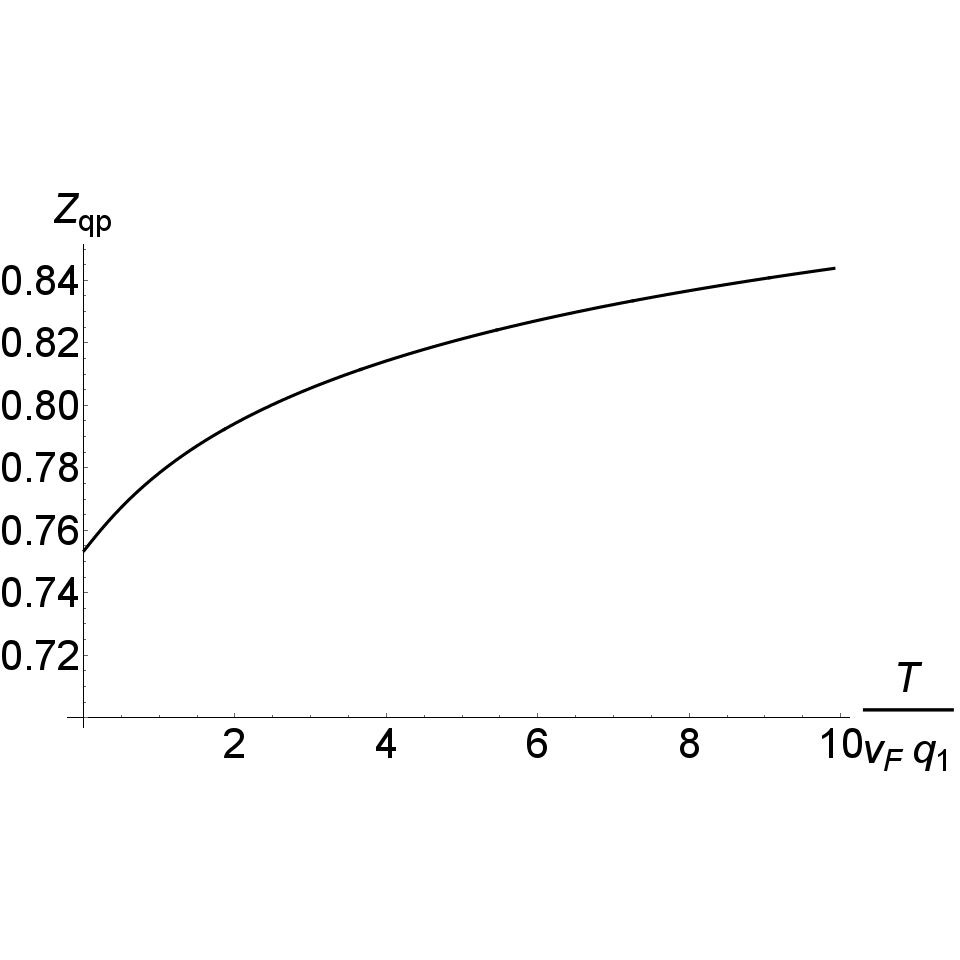}}
   \caption{Self energy of a finite system. In an infinite system the self energy has a non-analytic behavior at low temperatures, $\Sigma(\w) \sim \w^{2/3}$. In a finite system the nonanalytic behavior is cut off at a scale of $\w \sim \vf q_1$. The left panel depicts $\Sigma(\w)$ for a system of size $L = 20 a$. The dashed grey lines are guides to the eye of $\w, \w^{2/3}$. Note how the FL type behavior extends over a numerically large region near $\vf q_1$. The right panel depicts the quasiparticle residue as obtained from Eq. \eqref{eq:GF-tau} [26].
 \label{fig:sig-zqp}}
\end{figure}

Eq. \eqref{eq:sigma-final}
 is the explicit version of Eq. \eqref{eq:sigma-intro}.  We plot $\Sigma (\omega_m)$ along with QC and FL asymptotics ($\omega^{2/3}$ and $\omega$, respectively) in Fig. \ref{fig:sig-pl}. We see that in a finite-size system $\Sigma (\omega_m)$ preserves a FL form up to large $\omega_m/v_F q_1$.

An additional popular probe in QMC is the Green's function on the Fermi surface along the imaginary time axis \cite{Trivedi1995},
\begin{equation}
  \label{eq:GF-tau}
 G(\tau,\vfv) = T \sum_{\w_n} \frac{e^{i\w_n \tau}}{i\w_n - \Sigma(\w_n)}
\end{equation}
In a FL, $G(\tau=T/2)
=
Z_{qp}/2$. For  $\Sigma \sim \w_n^{2/3}$  $G(\tau=T/2)= T^{1/3}$, indicating that the quasiparticle residue vanishes at $T=0$
(the actual power is $T^{1/2}$ due to special form of the self-energy at the first Matsubara frequency \cite{Chubukov2012}).
 Because the sum
 is dominated by the  terms with $n = O(1)$, the finite size behavior of $\Sigma$ is important. Plugging parameters extracted from the data of Ref. \cite{Schattner2016} into Eq. \eqref{eq:sigma-final} and
   substituting into Eq. \eqref{eq:GF-tau}, we obtain $Z_{qp}$, which decreases as a function of $T$, but still approaches a finite value at $T \to 0$
      (see Fig. \ref{fig:zqp-pl} ). In this limit, our calculation yields $Z_{qp} = 0.76$. Ref. \cite{Schattner2016}
      found a very similar $Z_{qp} = 0.75-0.85$ in the low temperature regime
       \footnote{In order to compare our analytic expressions in Eqs. \eqref{eq:free-prop-elec}, \eqref{eq:pi-1}, \eqref{eq:sigma-final} with the QMC data, we used the average value of the Fermi vector and velocity $\kf, \vf$ for a cubic lattice. We extracted the value of $\gamma$ by comparing the values of $\Pi(0,0)$ at $g = 0$ and finite $g$. We extracted $c^2$ by fitting the data for $q=0$ to a quadratic form. The value of $\chi_0$ was taken from Ref. \cite{Schattner2016}.}.

{\bf {\it Discussion.}}~~~
We showed that finite size
 effects modify
  the
  low energy properties of the particle-hole polarization bubble and the fermionic self energy near a $q=0$ QCP.  We found three effects: i) the slope of the  frequency dependence of $\Pi (q, \Omega_m)$  changes from its universal $1/q$ form to almost $q-$independent, ii) the bosonic mass gets a $q$ dependent correction; iii) the  non-analyticity of the electronic self energy is cut off below a certain frequency.
    In a real finite-size system the strength of (i) and (ii) is actually a bit smaller than in our
      model,
      where the polarization appears to vanish at $q = q_1$ for all $\Omega_m$. In practice, there will always be a residual polarisation coming from a) broadening due to finite self energy, and b) the irregularity of the separation between filled/empty states due to
     a discrete structure of the
     FS in a
     finite system. Both these features can be seen just by studying the left panel of Fig. \ref{fig:facades}.
       A translational symmetry breaking inherent in any finite system
        also induces broadening. Nevertheless our results do capture the main features observed in QMC studies.

Our results can also be applied
to magnetic conducting nanoparticles  near a  near a ferromagnetic/paramagnetic QCP.
 Magnetic nanoparticles have attracted attention in recent years due to biomedical and other applications. One implication is that the finite-size correction to the bosonic mass  will introduce strong wavelength-dependent hysteresis in the region where $q (T-T_c) \sim T_c /L $. Another is that for $|T - T_c| \leq T_c a/L$ the magnetic susceptibility saturates and the resistance obeys its Fermi liquid $T^2$ behavior. In a recent work \cite{Swain2015}, magnetic nanoparticles of composition Pd$_{1-x}$Ni$_{x}$ were tuned across the transition, as evidenced by their magnetic response. However, the resistance remained Fermi-liquid like. In a follow-up analysis \cite{Swain2016} of Ni$_{1-x}$V$_{x}$, a decrease in the slope of $\chi(T)$ was observed, at $T \sim 5 - 10 K$, even at the QCP ($T_c \sim 0$).
 In these systems, $a \simeq 3.5 \mathring{A}$ and $L \simeq 20-40 \mbox{nm}$. Combining this with bulk nickel's
 Curie temperature $T_c \sim 600 K$ (to be distinguished from the nanoparticle $T_c \sim 0$ at the QCP) we find a saturation temperature of $ |T-T_c| = T_c a/L  = 5.5-11 K$, in remarkable agreement with experimental results. We leave further analysis of such systems to future work.

\acknowledgements
We thank E. Berg, S. Lederer, S. Kivelson, Y. Schattner, D. Chowdhury, R. Fernandes, X. Wang and S. K. Srivastava for useful discussions.   This work was supported by the NSF DMR-1523036.

\bibliography{QCP,QCP_added}

\begin{thebibliography}{29}
\bibitem [{ome()}]{omega23}%
  \BibitemOpen
  \href@noop {} {}\bibinfo {note} {We leave proof of this statement to a future
  publication.}\BibitemShut {Stop}%
\end{thebibliography}

\clearpage
\onecolumngrid
\appendix
\section{Supplemetary material}
\label{sec:suppl-mater}

\subsection{Derivation of eq. \eqref{eq:h-i-cont}}
\label{sec:derivation-h-i-cont}

In this section we derive the form of Eqs.  \eqref{eq:h-i-cont} + \eqref{eq:cutoff-cont} for the interaction in a finite system at the continuous limit. We do this via the Poisson summation formula:

\begin{align}
  \label{eq:supp-hi-finite}
  H_I &= \sum_{\q,\k} f_{\k} \phi_{\q} \pc[\k+\frac{\q}{2}] \pa[\k-\frac{\q}{2}] \nn \\
      &= \sum_{\q}\int_{\mathcal{C}}d^2p ~ f_{\p} \phi_{\q} \pc[\p+\frac{\q}{2}] \pa[\p-\frac{\q}{2}]\sum_{\k}\delta^{(2)}(\p - \k).
\end{align}
Here, $\mathcal{C}$ is a region of $p-$ space that covers all the $\delta$ functions, i.e. it is the first BZ except for an arbitrarily chosen finite strip around the Fermi surface. The strip configuration depends on the dispersion and chemical potential, but for simplicity we choose this strip to be of constant width $W = \alpha q_1$, where $0 < \alpha < 2$, and ignore any further geometric details. Expanding the $\delta-$ functions and performing the lattice sum in the usual way we obtain:
\begin{align}
  \label{eq:supp-hi-finite-2}
  H_I
  &= a^2\sum_{\q}\sum_{\bv{x}}\int_{\mathcal{C}}d^2p e^{-i N \bv{x} \cdot \p} ~ f_{\p} \phi_{\q} \pc[\p+\frac{\q}{2}] \pa[\p-\frac{\q}{2}].
\end{align}
Here, the $\bv{x}$ sum is over the lattice points of an infinite system, and the $N=L/a$ factor in the exponent comes from the fact that we are summing over the reciprocal space to $k-$ space, which has a lattice constant of $2\pi/L$. Let us show that all $\bv{x} \neq 0$ terms are small. To do so we assume that the integrand is slowly changing and treat it as a constant. For simplicity we also approximate the Fermi surface as a square of side length $2\kf$. We keep only those terms in the $\bv{x}$ summation that are parallel to the $x$ or $y$ axes, as all other terms oscillate rapidly when integrating over the surface. Then integrating Eq. \eqref{eq:supp-hi-finite-2} gives
\begin{equation}
 \delta H_I \sim \sum_{x=a n}\frac{a \kf \sin(\alpha n \pi)}{N n}
\end{equation}
where we have assumed that integrals along the $x$ and $y$ axes give similar values.
We see that $\alpha$ provides a modulation that cuts off higher $n$ terms, so that the total error is also of order $1/N$. We choose $\alpha = 1$, which gives a naive minimization of the errors, and keep only the $\bv{x} = 0$ term, yielding eqs. \eqref{eq:h-i-cont}+\eqref{eq:cutoff-cont}. Neglecting variations of the integrand is justified as long as
\begin{equation}
  \label{eq:supp-approx-validity}
   \frac{q}{q_1} \ll L/a  \Rightarrow q \ll \frac{1}{a}.
\end{equation}

\subsection{Evaluation of the polarization and self energy}
\label{sec:eval-polar-self}

In this section we evaluate the one-loop polarization and self-energy in our approximate model for a finite systems and derive Eqs. \eqref{eq:pi-1} + \eqref{eq:sigma-final}. The starting point is the one-loop bubble
\begin{align}
  \label{eq:pi-def}
  \Pi(\q, i\W) &= g^2\chi_0 T\sum_{\w = (2n+1)\pi T} \int \frac{d^2k}{(2\pi)^2} \left[i\w - \ve_{\k}\right]^{-1}\left[i(\w+\W) - \ve_{\k+\q}\right]^{-1}\Phi(\k,\q) f_{\k}^2\nn \\
               &\simeq \frac{g^2\chi_0 m}{(2\pi)^2}\int d\xi d\theta \frac{n_F(\xi +\frac{1}{2}\vf q\cos\theta)-n_F(\xi-\frac{1}{2}\vf q\cos\theta)}{i\W - \vf q \cos\theta}\times f^2(\theta)\nn \\
  &\qquad\quad \times\Theta\left(\vf q |\cos\theta| - \vf q_1\right)\Theta\left(\vf q \cos\theta - \vf q_1 - 2|\xi|\right).
\end{align}
Here, we have summed over the imaginary frequencies, and expanded the fermion energy near the FS. $n_F$ is the Fermi-Dirac distribution. Performing the $\xi$ integration and noting that the $\theta$ integral consists of two equal contributions from the ranges $(-\theta_\beta,\theta_\beta),(\pi-\theta_\beta,\pi+\theta_\beta)$  yields Eq. \eqref{eq:Pi-model}. The integral can be done analytically and yields Eq. \eqref{eq:pi-1}.

Next we compute the self energy, also in the bare one-loop approximation. One technical problem is to evaluate the polarisation in the regime $\alpha,\beta \ll 1, \W\sim \vf q_1$. The limits $\alpha,\beta \ll 1, \beta < \alpha$ and $\alpha \ll 1, \beta > \alpha$ do not commute, being linear and quadratic in $\alpha$ respectively. However, we will show that in this regime the $\W$ dependent term in the susceptibility can be neglected. We can therefore use only the expression for $\Pi$ in the regime $\beta < \alpha$, which was shown in the text right after Eq. \eqref{eq:pi-1}. We plug this expression into Eq. \eqref{eq:sigma-def} for $\Sigma$, and perform the angular integration obtaining,
\begin{align}
  \label{eq:supp-se-1}
  i\Sigma(\k,i\w) &= \frac{g^2\chi_0}{(2\pi)^2}\int d\W q dq \left[m^2 + q^2 + \gamma \frac{|\W|}{\vf q} + \gamma \log\frac{2\vf q}{|\W|}\frac{q_1}{q}\right]^{-1} \frac{\mbox{sign}(\w+\W)}{\vf q\sqrt{1+|\w+\W|/\vf q}} \nn \\
  &\simeq \frac{2g^2\chi_0}{(2\pi)^2\vf}\mbox{sign}(\w)\int_0^{|\w|}d\W\int_0^{\infty}dq \frac{q}{m^2q + q^3 + \gamma \frac{\W}{\vf} + \gamma \log\frac{2\vf q}{\W}q_1}.
\end{align}

For small $q$, the mass term is negligible as long as $m a \ll \sqrt{a/L}$, and we drop it. The denominator has three regions where it may be small: $q\sim 0,\W\sim q, q^3 \sim\W,q_1$. It is easy to see that in the first two regions the integrand is respectively small or finite. We can therefore assume that the logarithmic term is large and slowly varying, and treat it as a constant to be determined at the end of the calculation. A similar analysis confirms we can ignore the regime $\alpha < \beta$. We perform the $\W$ integration and obtain,
\begin{align}
  \label{eq:supp-se-2}
  i\Sigma &= \frac{2 g^2\chi_0}{\gamma(2\pi)^2}\mbox{sign}(\w) \int_0^{\infty} dq~q \log \left[1 + \frac{\gamma \w/\vf}{q^3+\gamma q_1 \log\alpha_L}\right] \nn \\
  &= \frac{2 g^2\chi_0}{\gamma(2\pi)^2}\mbox{sign}(\w)\left(\frac{\gamma\w}{\vf}\right)^{2/3}\int_0^{\infty}dq~q \log\left[1 + \frac{1}{q^3 + \frac{\vf q_1}{\w}\log\alpha_L}\right],
\end{align}
which gives Eq. \eqref{eq:sigma-final}.

\end{document}